\documentclass[12pt]{article}

\usepackage{graphicx}
\usepackage{multirow}

\newcommand{\beq}{\begin{equation}}
\newcommand{\eeq}{\end{equation}}
\newcommand{\beqa}{\begin{eqnarray}}
\newcommand{\eeqa}{\end{eqnarray}}
\newcommand{\beqar}{\begin{eqnarray*}}
\newcommand{\eeqar}{\end{eqnarray*}}

\begin{tiny}

\end{tiny} 

\begin{document}

\thispagestyle{empty}

\vspace{32pt}
\begin{center}

\textbf{\Large Magnetic fields and cosmic ray anisotropies\\ 
at TeV energies}

\vspace{40pt}

Eduardo Battaner, Joaqu\'\i n Castellano, Manuel Masip
\vspace{12pt}

\textit{Departamento de F{\'\i}sica Te\'orica y del Cosmos}\\
\textit{Universidad de Granada, E-18071 Granada, Spain}\\
\vspace{16pt}
\texttt{battaner@ugr.es, jcastell@correo.ugr.es, masip@ugr.es}

\end{center}

\vspace{30pt}

\date{\today}

\begin{abstract}

Several cosmic-ray observatories have provided 
a high accuracy map of the sky at TeV--PeV energies.
The data reveals an $O(0.1\%)$ deficit from north galactic 
directions that peaks at 10 TeV and then evolves with the energy,
together with other anisotropies at
smaller angular scales. Using Boltzmann equation 
we derive expressions for the cosmic-ray flux that fit
these features. The anisotropies depend on the 
local interstellar magnetic field 
$\vec B_{IS}$, on the average galactic field 
$\vec B_{R}$ in our vicinity, and on correlations
between fluctuating quantities. We show that the 
initial dipole anisotropy along $\vec B_{IS}$ can be modulated
by changes in the global cosmic ray flow, and that 
a variation in the dipole direction would imply a given
radius of coherence for the local $\vec B_{IS}$.
We also show that small and 
medium-scale anisotropies may appear when the 
full-sky anisotropy finds a field configuration 
acting as a magnetic lens.

\end{abstract}


\newpage

\section{Introduction}

We observe charged cosmic rays (protons and atomic nuclei)
with energies of up to $10^{11}$ GeV. Although in their way to the Earth 
these particles lose directionality, they carry
important information about their sources and about the 
environment where they have propagated. For example, the
observation that Boron is more frequent in cosmic
rays (CRs) than in the solar system suggests that it is produced
when heavier nuclei {\it break} in their
way to the Earth, implying that they cross an average
depth of 10 g/cm$^2$ of interstellar (IS) baryonic matter 
in their trajectory from the dominant 
sources \cite{Reeves:1970zz}.

At TeV energies magnetic fields trap charged CRs in the 
IS medium, and their transport is usually modeled 
by a diffusion equation \cite{shalchi}. One expects that 
the CR gas propagates along the parallel and the perpendicular
directions to the background field $\vec B$ 
with different diffusion coefficients, 
scattering with the magnetic turbulences
$\delta \vec B$ in the plasma. In particular, if we define the
Larmor radius  
\beq
r_L={E \over Q B c}=\left( {E\over 1\; {\rm TeV}} \right)
\left( { 1 \; \mu {\rm G} \over B} \right)
\left( { e \over Q}\right)\, 1.1\times 10^{-3} \;{\rm pc}
\label{larmor}
\eeq
a CR will predominantly be diffused by magnetic irregularities
of wave number $k\approx 1/r_L$.
In a first approximation, one may picture its trajectory
as an helix along $\vec B$ of radius $r_L\sqrt{1-\mu^2}$ and velocity
\beq
v_{\parallel}=c \mu\,, \;\;\;v_{\perp}=c \sqrt{1-\mu^2}\,,
\eeq
with random changes in $v_{\parallel}$ after 
a parallel mean free path $\lambda_\parallel$. Such change will 
also imply a variation in the field line trapping the CR, 
{\it i.e.}, $\lambda_\perp\approx r_L$. 

A diffusion equation admits a multipole expansion \cite{jones90}
with isotropy at order zero and a dipole along the gradient
direction at first order.
However, this information is deduced {\it a posteriori}, 
as in a diffusion equation 
the momenta of the gas particles have been averaged. 
Boltzmann equation, instead,
gives the evolution in phase space 
of the statistical distribution function $f(\vec r,\vec p; t)$
(density of particles at $\vec r$ with momentum $\vec p$), 
providing a microscopic description of the fluid
\cite{Battaner:2009zf,Ahlers:2013ima}.
It is easy to see that when we measure
the CR differential flux $F(\vec u, E;t)$ 
(number of particles crossing the unit 
area from a given direction $\vec u$ per unit 
solid angle, energy and time) we can directly {\it read}
the distribution function:
\beq
f(\vec r_{Earth}, -\!\, {E\over c}\vec u ; t) =
{c^2\over E^2}\, F(\vec u, E;t)\,,
\label{flux0}
\eeq
where we have taken the relativistic limit with $E=cp$. 
Therefore, it is
interesting to explore how the appearance of anisotropies
may be explained with Boltzmann equation, specially in 
an environment with regular magnetic fields at different scales
(see below).

In this article we will attempt a description of several large 
and medium scale anisotropies observed in the CR
flux by several experiments. 
The combined  results from TIBET \cite{Amenomori:2005dy}, 
MILAGRO \cite{Abdo:2008aw}, ARGO-YBF \cite{ARGO-YBJ:2013gya},
SuperKamiokande \cite{Guillian:2005wp}, 
ANTARES \cite{Mangano:2009ee},
IceCube \cite{Aartsen:2012ma,Santander:2013gma}
and HAWC \cite{Abeysekara:2013rka} provide
a picture of the whole sky at different energies. The data 
reveals that the almost perfect isotropy is broken by 
a $O(10^{-3})$ dipole-like feature that appears at 1 TeV and
evolves with the energy, together with other 
irregularities at lower angular scales \cite{Zotov:2012se,Iuppa:2013gha}. 

It seems clear that the  direction of the local IS
magnetic field $\vec B_{IS}$ 
should be a key ingredient in the explanation of these 
anisotropies \cite{schwadron2014}. 
Voyager data \cite{Ratkiewicz2008} on the heliospheric boundary 
provides an estimate for the direction of $\vec B_{IS}$:
\beq
\ell_B=217^o\pm 14^o\,;\;\;\;\;
b_B=-49^o\pm 8^o\,
\label{b1}
\eeq
(in galactic
coordinates), whereas IS atom measurements with IBEX 
\cite{Frisch:2012zj} imply
\beq
\ell_B=210.5^o\pm 2.6^o\,;\;\;\;\;
b_B=-57.1^o\pm 1.0^o\,.
\label{b2}
\eeq
Although the region of coherence of such field is unknown
(it could vary from 0.01 to 10 pc), it is much larger than
the gyroradius of a TeV cosmic ray (in Eq.~(\ref{larmor})).
At even larger distances (above 10 pc) the average magnetic 
field $\vec B_R$ can be measured using a variety of methods 
\cite{beck2005,wiel2005,Han:2009ts,edu2009,RuizGranados:2010hf}: 
polarized thermal dust emission from clouds, Zeeman splitting of
lines or Faraday rotation of polarized galactic and
extragalactic sources, among
others. The data seems to
reveal
a $B_R\approx 3$ $\mu$G field pointing clockwise 
($\ell_B\approx 90^o$) in the galactic plane 
($b_B\approx 0^o$) \cite{Han:2009ts}. We will also study the role that
this global magnetic field plays in the explanation of the
CR anisotropy.

In Section 2 we start by discussing the anisotropy 
that would be expected for a single CR 
source and an isotropic propagation. We then  
use Boltzmann equation to analyse how the anisotropy 
is deformed by the presence of a magnetic field
$\vec B_{IS}$. We will assume that the 
dominant CR sources are beyond the region of coherence
of $\vec B_{IS}$ and that their effect is captured by 
boundary conditions. 
In Section 3 we 
review the trajectory of CRs
in the absence of turbulences. In particular, 
we study the {\it image} of a point-like source and 
show that there are many trajectories connecting the source
with a given observer. Such study will be necessary to understand 
the appearence of small and medium-scale anisotropies. 
In sections 4 and 5, respectively, we analize the data 
and sumarize our results.

\section{Large-scale anisotropies}

Let us first consider the simplest flux: 
a CR gas from a pointlike source $S$ propagating 
through a turbulent but homogeneous and isotropic medium. Such  
medium would correspond to the absence of regular 
magnetic field $\vec B_{IS}$ (or to the presence of a 
field weaker than the 
fluctuations $\delta \vec B$ of wave number $k\approx 1/r_L$), 
and it implies the same diffusion coefficient $\kappa$ 
in all directions. 
The trajectories will define in this case a three-dimensional
random walk of step $\lambda=3 \kappa/c$. The mean displacement $D$ 
from the source that a particle reaches after a (large) time
$t$ is then \cite{shalchi}
\beq
D = \sqrt{ 2 \kappa t}\,.
\eeq
The expression above implies that the radial velocity of the gas  
(we call it the {\it CR flow}, since an observer moving at
that velocity would observe an isotropic flux) 
will decrease like $1/\sqrt{t}$ with the distance $D$ from $S$:
\beq
v_{\rm gas}\approx \sqrt{2\kappa\over t + 2\kappa/c^2}
= c\left( {c^2 D^2\over 4 \kappa^2} + 1 \right)^{-1/2}\,.
\eeq
The relative difference between the CR flux 
going away or towards the source (the forward-backward
asymmetry $A^{FB}$) can then be estimated as the ratio
\beq
A^{FB}\approx {v_{\rm gas}\over c} \approx
{2\kappa \over c D}\,.
\eeq
This means that the point-like source will introduce an anisotropy
in the CR flux proportional to $1/D$ and to 
$\lambda$. Basically, it 
is a dipole anisotropy with the excess pointing towards $S$:
\beq
F(\vec u)= F_0 \, (1 +  \vec u \cdot \vec d)\,,
\label{flux1}
\eeq
where
\beq
\vec d = {A^{FB}\over 2 \pi} \,\vec u_S\,.
\eeq

When there are several sources $S_i$, it is straightforward 
to show that the addition of the corresponding 
dipole anisotropies $\vec d_i$ 
gives {\it another} dipole $\vec d$ \cite{Giacinti:2011mz}:
\beq
\vec d = {\sum_i F_0^{(i)} \,\vec d_{i}\over 
\sum_j F_0^{(j)}}\,.
\eeq

In summary, for an isotropic CR propagation 
we may expect a dipole anisotropy pointing towards the
average CR source \cite{Pohl:2012}, with its intensity 
inversely proportional  to the distance to these sources and 
proportional to the mean free path between collisions.
Notice, however, that the presence of a regular magnetic
field $\vec B_{IS}$  will introduce
an asymmetry between the parallel and the perpendicular 
diffusion coefficients ($\kappa_\parallel$ and $\kappa_{\perp}$) 
that will change this result.

To find out how, let us assume a local $\vec B_{IS}$ 
coherent over distances $R_{IS}\gg r_L,\lambda_{\parallel}$, with
$\lambda_{\parallel}=3 \kappa_{\parallel}/c$. We will treat the CRs
(protons of energy between $1$ and $1000$ TeV) as a fluid
that only interacts with the magnetic fields. 
To obtain the average CR anisotropy in our vicinity 
we will separate the magnetic field and the distribution function 
into a regular plus a turbulent component, 
\beqa
f & \rightarrow & \bar f +\delta f\;,\cr
\vec B & \rightarrow & \vec B_{IS} +\delta \vec B\;.
\eeqa
with 
\beq
\langle \,\delta \vec B\, \rangle = \langle\, \delta f\, \rangle = 0\;,
\eeq
and we will  
{\it average} Boltzmann equation over nearby points.
Given the relatively small distance and time scales, 
we will take stationary and homogeneous 
magnetic field $\vec B_{IS}$ and distribution function $\bar f$. 
We are then assuming that the CR sources are far enough so that 
the spatial gradient $\nabla_{\! r} \bar f$ is negligible ({\it i.e.},
smaller than $\delta f/R_{IS}$), that the changes in $f$ occur on 
time scales much larger than the period of data taking (the movement
of the Earth around the Sun introduces 
irregularities of order $10^{-4}$, {\it i.e.}, a $10\%$ correction
to the large scale anisotropy under consideration),
and we ignore energy loss or collisions with IS matter.
For a fixed CR energy, $\bar f$ must satisfy Boltzmann equation:
\beq
\vec F \cdot \nabla_{\! u}\, \bar f(\vec u)
= e\; (\vec u \times \vec B) \cdot 
\nabla_{\! u}\, \bar f(\vec u)=0\;,
\eeq
where $\vec u=\vec p/p$ and $\vec p$ is the momentum of the CR.
The equation above can also be written 
\beq
\vec{u} \cdot \left( \vec{B} \times \nabla_{\! u}\, \bar f \right)=0\;,
\eeq
which admits the generic solution 
\beq
\bar f(\vec u)= \bar f (\vec u \cdot \vec u_B)\,.
\eeq
Any stationary and homogeneous solution must then be 
a function with symmetry 
around the axis of the magnetic field:
$\vec B_{IS}$ will {\it isotropize} the flux in the directions 
orthogonal to its axis. 
In particular, these solutions may
accommodate a dipole along $\vec u_B$,
\beq
\bar f(\vec u)=  f_0 \, \left( 1 -  
\vec u \cdot \vec d\right) \,,
\eeq
with 
\beq
\vec d = {A^{FB}\over 2 \pi} \,\vec u_B\,.
\eeq
This distribution function will define (see Eq.~(\ref{flux0})) 
the dipolar flux in (\ref{flux1}) with $\vec u_S\to \vec u_B$
and $\,F_0=f_0 (E/c)^2$,
{\it i.e.}, it is $\vec B_{IS}$ (and not the position of the sources)
what fixes the direction of the CR flow in our frame, 
defined\footnote{Notice that an observer moving at $\vec v_0$ will see
no net flux and complete isotropy. This velocity 
may coincide or not (for example, due to
an asymmetry in the location of CR sources) with the velocity of 
the local plasma wind.} as
\beq
\vec v_0/c= {1\over N}
\int {\rm d}\Omega \; f(\vec u)\; \vec u
\label{v0}
\eeq
with 
$\;N = \int {\rm d}\Omega \; f(\vec u)\,$.

\begin{figure}[t]
\begin{center}
\includegraphics[width=7.cm]{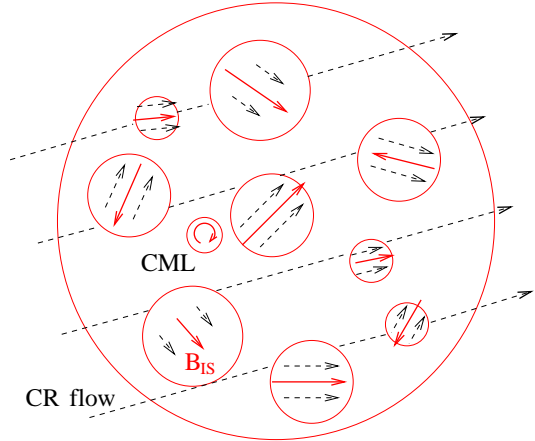}
\label{wind}
\vspace{-0.3cm}
\caption{
$\vec B_{IS}$ (within coherence cells of 0.1--10 pc) and cosmic ray
flow. CML indicates a cosmic magnetic lens \cite{Battaner:2010bd}.
}
\end{center}
\end{figure}
The (forward or backward) direction along $\vec B_{IS}$ 
and the intensity of this dipole anisotropy 
will depend on boundary conditions that, in turn, 
will reflect the average direction
of the CR flow \cite{Biermann:2012tc,Qu:2011fe}
at larger scales. 

In Fig.~1
we plot a scheme of the CR flow within different cells that
contain a regular $\vec B_{IS}$. This field may change randomly from
cell to cell, although it has an average value
$\vec B_R$ at kpc scales \cite{Han:2009ts}. Our result above
has been obtained for an observer at rest within our local IS medium.
The CR anisotropy in each cell (which may have a velocity 
relative to us) will then follow the $\vec B_{IS}$ magnetic lines, with 
a forward or backwards direction depending on the 
projection of the global (average) flow $\vec d_R$ along $\vec B_{IS}$.
In particular, notice that $\vec d\approx 0$ for
a $\vec B_{IS}$ orthogonal to $\vec d_R$. 
Notice also that we are neglecting the 
velocity $v\approx 23$ km/s \cite{McComas:2012}
of the Sun relative to our cell of 
local IS medium and the velocity $v'\approx 30$ km/s
of the Earth around the Sun. These movements imply
Compton-Getting \cite{C&G} irregularities
--in the sense that they are caused by the velocity of the
observer-- of order $v/c = O(10^{-4})$, introducing $10\%$ 
corrections to the dominant anisotropy along $\vec B_{IS}$.

An important question would then be what to expect for the average
CR flow. Is it $\vec d_R =\langle\, \vec d\, \rangle$
a dipole along the direction of the average magnetic field
$\vec B_R = \langle\, \vec B_{IS}\, \rangle$? To unswer this question
we again separate the magnetic field and the distribution function 
into a regular plus a fluctuating component, but now we 
average Boltzmann equation over larger distances, which will  
include different (nearby) cells:
\beqa
f & \rightarrow & f_R +\delta f\;,\cr
\vec B & \rightarrow & \vec B_R +\delta \vec B\;.
\eeqa
Although $\delta \vec B$ and $\delta f$ vary randomly
from one cell to another, there may be correlations between
both turbulent components (i.e., their relative value in each
cell is not random). We will assume
\beqa
\langle\, e\; (\vec u \times \delta \vec B) \cdot 
\nabla_{\! u}\; \delta f \,\rangle &=& 
e\; \vec u \cdot \langle\, \delta \vec B\times 
\nabla_{\! u}\; \delta f \rangle\cr
&=& e\; \vec u\cdot \vec T\,.
\eeqa
Boltzmann equation for the regular components is then 
\beq
\vec{u} \cdot \left( \vec{B_R} \times \nabla_{\! u}\; f_R \right)+
\vec{u} \cdot \vec T=0\,.
\label{boltzmann}
\eeq
We can
\begin{figure}[t]
\begin{center}
\includegraphics[width=5.cm]{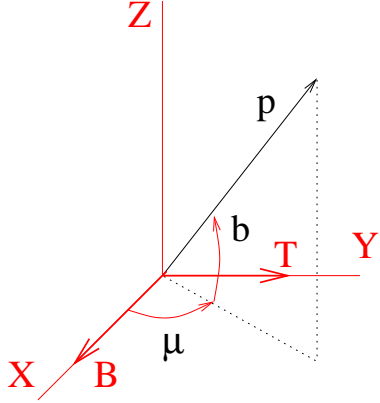}
\label{coord}
\vspace{-0.3cm}
\caption{
Coordinate system. 
}
\end{center}
\end{figure}
find consistent solutions when the correlation 
$\vec T$ is constant and orthogonal to $\vec B_R$.
We place the axes (see Fig.~\ref{coord}) so that $\vec B_R$ and
$\vec T$ go  
along the $X$ and the $Y$ axis, respectively, and we use the 
latitude $b$ and the 
longitude $\mu$  to label the direction $\vec u$ of a CR.
Taking $f_R(\vec u)=f_R(b,\mu)$ and
\beq
\nabla_{\! u}\; f_R = {\partial f_R\over \partial b} \;\vec{u}_b + 
 {1\over \cos b}\; {\partial f_R\over \partial \mu} \;\vec{u}_\mu  
\eeq
with 
\beq
\vec u_b=-\sin b\;\cos\mu \;\vec u_\phi-\sin b\;\sin \mu \;\vec u_r
+\cos b \;\vec u_z\; ; \;\;\;
\vec u_\mu= -\sin \mu \;\vec u_\phi + \cos \mu \;\vec u_r
\;,
\eeq
Boltzmann equation becomes
\beq
-\sin\mu\; {\partial f_R\over \partial b} +
\tan b\; \cos\mu\; {\partial f_R\over \partial \mu}
+{T\over B_R} \; \cos b \; \sin\mu  =0\;.
\label{sol0}
\eeq
This equation can be solved 
analytically:
\beq
f_R(b,\mu)= f_0 \left( 1+{T\over f_0 B_R}\;\sin b \right)+ 
\tilde f (\cos b\, \cos\mu) \;,
\eeq
with $f_0$ a constant and $\tilde f$ an arbitrary function of 
$\cos b\, \cos\mu$. We see that 
the first term is just a dipole orthogonal
to the plane defined by $\vec B_R$ and $\vec T$, whereas the second 
term may include a dipole along $\vec B_R$:
\beq
f_R(b,\mu)= f_0 \left( 1+t\, \sin b + s\,
 \cos b\, \cos\mu \right) \;,
\eeq
with $t=T/(f_0 B_R)$ and $s$ a constant depending on boundary
conditions. The CR flux that corresponds to this distribution
function (see Eq.~(\ref{flux0})) would be
\beq
F_R(\vec u)= F_0\, \left( 1+ \left( \vec d_t + \vec d_s \right)
\cdot \vec u  \right) \;,
\label{flux2}
\eeq
where
$\,F_0=f_0(E/c)^2\,$, $\,\vec d_t = - t\, \vec u_B\! \times\! \vec u_T\,$ and
$\,\vec d_s = - s\, \vec u_B\,$.
Eq.~(\ref{flux2}) expresses a key result: the global CR flow
$\vec d_R$
does not necessarily flow along the average magnetic field $\vec B_R$. 
There may appear a 
second dipole anisotropy orthogonal to $\vec B_R$ 
that, added to the first dipole, could favor any direction:
$\vec d_R =\vec d_t + \vec d_s$. Moreover, 
the turbulent correlation $\vec T$ 
defining this second dipole may evolve with the energy 
and vary its direction, which would
translate into a change in the global CR flow and 
then in the boundary conditions that determine
the dipole anistropy along $\vec B_{IS}$ 
described above. 

We would like to make some final observations concerning
the evolution of the anisotropy with the energy. 
For a standard 
Kolmogorov spectrum of magnetic turbulences
$\lambda_\parallel$ grows with the energy like 
$\approx E^{0.6}$ \cite{Swordy}, 
whereas $\lambda_\perp\approx r_L$ 
increases linearly with the CR energy. When the parallel
and the transverse mean 
free paths become similar the propagation becomes
isotropic and we should see the global
CR flow (see Fig.~\ref{wind}). This flow, in turn, should
reflect the velocity of our local IS plasma and the 
position and the intensity of the average CR source. 
Moreover, the
isotropic propagation would also be a sign that $r_L$
has reached a size similar to the 
region of coherence of $\vec B_{IS}$, since the 
fluctuations $\delta B$ of wave number $k\approx 1/r_L$ should
be $\delta B\approx B_{IS}$.

\section{Small and medium scale anisotropies}

A small scale anisotropy in the CR flux 
must be generated closer to the Earth 
\cite{Salvati:2010pq,Giacinti:2011mz},
at distances where the diffusive regime has 
not been fully established yet.
It is then necessary to study the image of a point-like 
CR source after crossing a constant magnetic field, without 
the magnetic turbulences that cause the
diffusion.

A particle of charge $Q$ and energy $E\gg mc^2$
in a constant field $\vec B_{IS} = B_{IS}\, \vec k$ will describe
a helix of angular frequency $\omega=B_{IS}cQ/E$ and radius 
$r(\mu)=c\sqrt{1-\mu^2}/\omega$. Choosing the coordinates
such that at $t=0$ the particle is at $S=(0,0,0)$ the trajectory reads
\beqa
x&=& - r(\mu)\, \sin \phi_0
+r(\mu)\,
\sin \left(\phi_0+\omega\, t\right)\nonumber\\
y&=&  r(\mu)\, \cos \phi_0
- r(\mu)\,
\cos \left(\phi_0+\omega \,t\right)\nonumber\\
z&=& c\, \mu \,t\,,
\eeqa
where $\mu=v_{\parallel}/c$ and $\phi_0$ is the inital angle
of $v_{\perp}$ with the $X$ axis:
\beqa
\dot x&=&c\,\sqrt{1-\mu^2}\,
\cos \left(\phi_0+\omega\, t\right)\nonumber\\
\dot y&=&c\,\sqrt{1-\mu^2}\,
\sin \left(\phi_0+\omega\, t\right)\nonumber\\
\dot z&=&c\, \mu \,.
\eeqa
\begin{figure}[t]
\begin{center}
\includegraphics[width=7.cm]{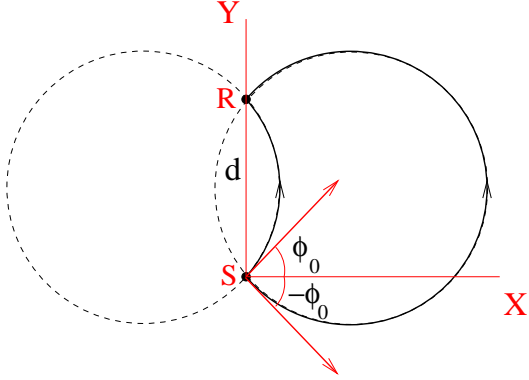}
\end{center}
\label{circles}
\vspace{-0.5cm}
\caption{
Trajectories between $S=(0,0,0)$ and $R=(0,d_\perp,0)$ for
$\vec B_{IS} = (0,0,B_{IS})$. 
}
\end{figure}
Let us consider all the trajectories connecting the source $S$
with an observer $R$ located at a transverse distance 
$d_\perp\le c/\omega=r_L$
and a paralel distance $d_\parallel\ge 0$. We can always rotate the axes so
that $R$ is at $(0,d_\perp,d_\parallel)$ 
and use the variables $(\mu,\phi_0,t)$ to 
solve $(x,y,z)=(0,d_\perp,d_\parallel)$. It turns out that there
is an infinite number of such trajectories, each one characterized
by an integer {\it winding} number $n\ge n_{\rm min}$, with
\beq
n_{\rm min}={\rm Integer} \left[ {d_\parallel\over \pi 
\sqrt{4 r_L^2-d_\perp^2}} \right]\,,
\eeq
and a (positive or negative) 
$\phi_0$ with $|\phi_0|\le \pi/2$. 
To see this it is instructive
to first consider the case with $d_\parallel=0$ (in Fig.~\ref{circles}), 
{\it i.e.}, with $S$ and $R$ in a plane orthogonal to $\vec B_{IS}$. 
The trajectories in this case have $\mu=0$, $\phi_0^-=-\phi_0^+$ 
and will reach $R$ after an arbitrary number $n$ of turns 
around the left or the right circles in Fig.~\ref{circles}. Notice
that higher values of $n$ correspond to 
longer trajectories, which will provide fainter images of $S$ 
(the flux scales like $1/L^2$). 
Adding a distance $d_\parallel$ along the
$\vec B_{IS}$ direction the trajectories will require a non-zero value of 
$\mu$ to reach $R$, with  $L= d_\parallel/ \mu$ their total length. Trajectories
with larger values of $\mu$ will be brighter, although this parameter
is bounded by the condition $r_L \sqrt{1-\mu^2} \ge d_\perp/2$. 

In Fig.~4
\begin{figure}[t]
\begin{center}
\begin{tabular}{c}
\includegraphics[width=6.5cm]{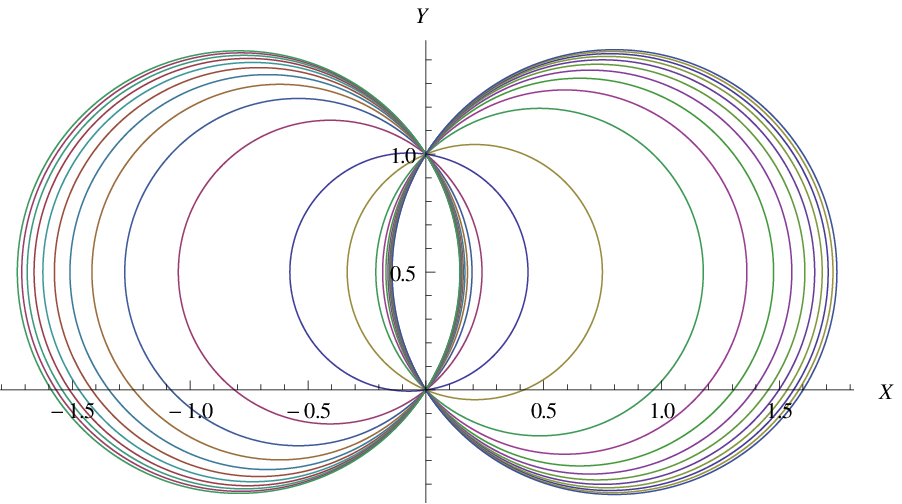} 
\end{tabular}
\hfil
\begin{tabular}{c}
\includegraphics[width=7.5cm]{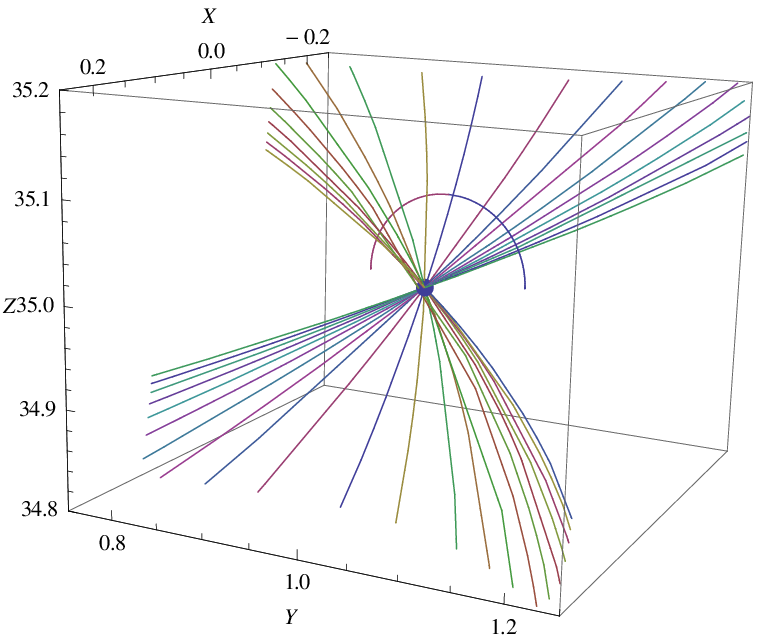}
\end{tabular}
\label{traj2}
\end{center}
\vspace{-1cm}
\caption{
Twenty shortest trajectories between $S=(0,0,0)$ and $R=(0,1,35)$ 
for $\vec B_{IS}=(0,0, B_{IS})$ and $r_L=1$: 
projection on the XY plane (left) and trajectories at $R$ (right).
In the limit $d_\parallel\gg d_\perp$ the source is {\it seen} at $R$ as 
a semi-conus of angle $\theta=\arccos d_\perp/(2r_L)$
with its axis along X and the limiting directions 
($\varphi=\pm \pi/2$) defining the XY plane.
}
\end{figure}
we plot several trajectories connecting $S$ with $R$ for a
large ($d_\parallel=35r_L$) longitudinal distance.
In the limit of very large $d_\parallel$ 
the trajectories arrive at $R$ defining a semi-conus
of directions of angle $\theta=\arccos d_\perp/(2r_L)$,  with 
$-\pi/2 < \varphi < \pi/2$ and the limiting 
directions ($\varphi=\pm \pi/2$) defining the plane orthogonal 
to $\vec B_{IS}$. It is easy to see that the trajectories
with direction $\varphi=0$ and maximum $\mu$ are shorter but less dense
than the ones in the extremes. As a consequence, the 
brightness (number of trajectories per unit length times their
flux) along the semicircle scales like 
\beq
{\cal B} = {\cal B}_0 \cos \left( \varphi+\pi/2 \right)\,.
\eeq
Notice also that each trajectory reaching $R$ corresponds
to a CR that left
the source $S$ at a different time, so the image at $R$ would be
the whole semicircle only for a constant and isotropic source.

Although from the previous analysis it is apparent that a nearby 
source could introduce small and medium scale anisotropies 
in the CR flux, we do not expect any sources
at distances below $1$ pc, which would probably introduce
too large anisotropies. We find, however, another plausible
mechanism for the generation of this type of  anisotropies.

In \cite{Battaner:2010bd} we have described the possible effects of a 
{\it cosmic magnetic lens}: a predominantly
toroidal field configuration that may appear with a variety of sizes 
and magnetic strength. As deduced from 
Liouville’s theorem\footnote{This theorem, first applied to 
cosmic rays moving inside a magnetic
field in \cite{Lemaître:1934}, 
implies that an observer 
following a trajectory will always
observe the same differential flux (or intensity, particles 
per unit area and solid angle) along
the direction defined by that trajectory.}, 
an isotropic and homogeneous flux will never become
anisotropic due to the action of a magnetic field.
However, the large-scale dipole anisotropy discussed
in the previous section could cross a nearby field
configuration acting as a magnetic lens
and imply point-like anisotropies of the same order. 
The lens would then become equivalent to
a faint source of CRs but be otherwise
invisible, since it does not produce nor deflect the light.

\section{Comparison with the data}

SuperKamiokande, TIBET, ARGO-YBF, ANTARES, MILAGRO and, 
more recently, HAWC
have been able to distinguish from the northern sky an $O(0.1\%)$ 
large-scale anisotropy in the flux of 1--10 TeV CRs.
IceTop and IceCube have observed 
with also very high statistics up to the PeV scale 
from the South Pole. In Table~1, we give an estimate of the
results obtained in these experiments. 
It is remarkable that all the observations seem consistent
with each other, although
the higher energies seen at IceTop have not been 
accessible to the previous experiments nor to HAWC yet.
One should notice that each experiment can access all the right
ascensions ($\alpha$) but only a limited region 
of declinations ({\it e.g.}, $-90^o< \delta < -25^o$
in IceCube). It becomes then difficult to estimate 
whether the excess and the deffect in the flux are opposite 
to each other ($\alpha\to \alpha\pm 180^o$, $\delta\to -\delta$)
and define a dipole.
Actually, in most experiments the region of maximum excess 
or maximum deffect is found at the limiting declinations that are 
accessible, suggesting that the {\it real} maximum is out
of reach. If that is the case, the non-accessible pole 
will introduce a relatively less intense and broader
anisotropy than the pole that can be seen 
by the experiment.

\begin{table}[t]
\begin{center}
\begin{tabular}{|c|c|c|c|c|c|}
\hline
\multirow{2}{*}{Hemisphere} & \multirow{2}{*}{Experiment} 
& \multirow{2}{*}{$\langle E\rangle$ [TeV]} & 
\multicolumn{2}{c|}{Deficit Position} 
& \multirow{2}{*}{Amplitude} \\ \cline{4-5}
&  &  & R.A. [deg] & Decl. [deg] &       \\ \hline
\multirow{7}{*}{North}      & ARGO                    
& 3.6                               & 170 to 210          
& $-10$ to $30$           & $3 \cdot 10 ^{-3}$                \\ \cline{2-6}
                            & MILAGRO                 
& 6                              & 180 to 220          
& $-10$ to 0          & $3 \cdot 10 ^{-3}$             \\ \cline{2-6}
                            & TIBET                  
& 6.2                              & 170 to 210          
& $-10$ to $20$          & $3 \cdot 10 ^{-3}$                \\ \cline{2-6}
                            & ARGO                  
& 24                              & 150 to 190          
& $-10$ to $30$          & $1 \cdot 10 ^{-3}$                \\ \cline{2-6}
                            & TIBET                    
& 300                           & -                
& -                  & $ < 1 \cdot 10 ^{-3}$             \\  \hline
\multirow{4}{*}{South}      & ICECUBE              
& 20                             & 190 to 240          
& $-30$ to $-60$         & $8 \cdot 10 ^{-4}$              \\ \cline{2-6}
                            & ICECUBE               
& 400                           & 40 to 100            
& $-15$ to $-45$         & $7 \cdot 10 ^{-4}$              \\ \cline{2-6}
                            & ICETOP                 
& 400                           & 70 to 110            &    $-15$ to $-45$      
& $1.6 \cdot 10 ^{-3}$                  \\ \cline{2-6}
                            & ICETOP                   
& 2000                         & 50 to 125            
&     $-25$ to $-55$    & $3 \cdot 10 ^{-3}$                \\ \hline
\end{tabular}
\end{center}
\caption{ Summary of data on the large scale anisotropy 
obtained by several observatories:
ARGO \cite{DiSciascio:2011fk}; MILAGRO \cite{Abdo:2008aw}; 
TIBET \cite{Amenomori:2006bx}; ICECUBE \cite{Abbasi:2011zka}; 
ICETOP  \cite{Aartsen:2012ma}.
\label{table}}
\end{table}

The data can be summarized as follows.
At 1--20 TeV it reveals a dipole 
anisotropy that goes along  
\begin{figure}[t]
\begin{center}
\includegraphics[width=10.cm]{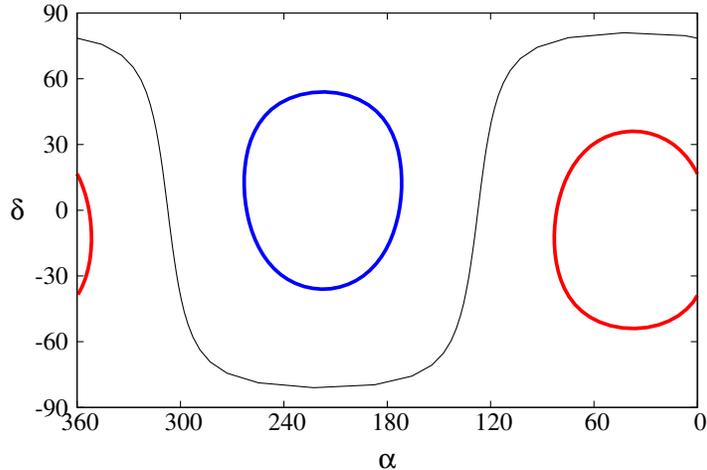}
\end{center}
\label{dipole}
\vspace{-0.3cm}
\caption{
Dipole anisotropy along $\vec B_{IS}$
for $\ell_B=180^o$ and $b_B=-60^o$ in equatorial coordinates
(right ascension and declination). The thin line indicates 
the magnetic equator, whereas thick lines define cones of 
angle $\pi/4$ along the magnetic axis.
}
\end{figure}
$\vec B_{IS}$. Taking all the data from the Northern observatories
and the low-energy IceCube results (based on the observation
of atmospheric muons) we estimate 
\beq
\ell_B=180^o\,;\;\;\;\;
b_B=-60^o\,, 
\label{b3}
\eeq
which is consistent with the values in Eqs.~(\ref{b1}-\ref{b2}).
In Fig.~5 we plot in equatorial coordinates 
two cones of angle $30^o$ with the axis along and opposite to the
direction of this $\vec B_{IS}$.
At higher energies the
observations from the South Pole indicate that the 
anisotropy weakens, becoming of order $10^{-4}$.
This result is supported by TIBET (see Table~1) and, especially,
by EAS-TOP \cite{Aglietta:2009mu}, which at $E\approx 100$ TeV 
is able to {\it see} the movement of the Earth around the 
Sun (an anisotropy of amplitude $2\times 10^{-4}$).
At even higher energies (around 400 TeV) both 
EAS-TOP and IceCube detect an increase in the amplitude
of the anisotropy and also a large change of phase, suggesting
a dipole almost opposite to the initial one. 
Finally, at 2 PeV \cite{Aartsen:2012ma}
the direction of the excess may have changed 
slightly towards the galactic center.

Our results in Section 2 provide a framework to 
interpret these observations. Above 1 TeV 
the effect of the heliosphere on the
CR trajectories is subleading, and the dominant magnetic field
is the $\vec B_{IS}$ in Eq.~(\ref{b3}).
The modulation above 10 TeV can then be explained if the
global CR flow varies its direction with the energy. 
In particular, if  
its component along $\vec B_{IS}$ changes sign 
at $\approx 100$ TeV. At these energies other effects, like
the Compton-Getting of order $10^{-4}$
due to the velocity of the Earth, become
relatively important. As the CR energy grows
the possible missalignment of the anisotropy
with $\vec B_{IS}$ would  indicate that $r_L\approx R_{IS}$, 
where $R_{IS}$ is the radius of
coherence of the local IS magnetic field (see Fig.~\ref{wind}).
For $B_{IS}\approx 3$ $\mu$G \cite{schwadron2014} and $E\approx 1$ PeV
we obtain $R_{IS}\approx 0.3$ pc.
The propagation becomes then isotropic \cite{Casse:2001be} and the
dipole anisotropy should follow the direction
of the global CR flow, which is driven by a correlation of 
turbulent quantities (at this scale, $\delta B=B_{IS}$ and the
fluctuation $\delta f$ generated by the movement of our local 
IS cell relative to our neighbours) and by the average magnetic field $B_R$.
The radius $R_{IS}\approx 0.3$ pc of our local IS plasma 
could also be related to the appearance
of the {\it knee} in the CR spectrum, as CRs of energy above 1 PeV
could not be trapped by $B_{IS}$ in our vicinity.

As for the low-scale anisotropies, we have described in
Section 3 how 
the image of the dipole through a cosmic magnetic lens 
may introduce irregularities. 
These could consist of 
pointlike and/or longitudinal structures 
similar to the ones discovered by some experiments:
the two regions observed
by TIBET \cite{Amenomori:2006bx} and MILAGRO \cite{Abdo:2008kr}
or the four regions (which include the two former regions) 
found by ARGO \cite{ARGO-YBJ:2013gya}. The lens could
{\it focus} the CR wind (see Fig.~1) and define anisotropies
of order $5\times 10^{-4}$, the amplitude that has been observed.
Notice also that, since the effect of the magnetic lens on 
a more energetic CR will be smaller \cite{Battaner:2010bd},
the irregularities will slightly change their position and
finally disappear when the energy grows. We think that 
region 2 in \cite{ARGO-YBJ:2013gya} --region $B$ in \cite{Abdo:2008kr}--
could be related to the effect that we described (region 1,
the most intense, seems linked to an effect of the 
heliotail \cite{Lazarian:2010sq}).

\section{Summary and discussion}

The appearence of anisotropies in the flux of charged CRs
provides information about the distribution of sources
and about the magnetic plasma where they
have propagated on their way to the Earth. The O($10^{-3}$) 
deficit from north galactic
regions discovered by TIBET and 
MILAGRO seems to follow the
direction of $\vec B_{IS}$. 
Using Boltzmann equation we have 
justified this observation and have shown that the CR
flow at more global scales may modulate this anisotropy, 
reducing its intensity and even inverting its direction
at higher energies. These features seem consistent with 
IceCube observations in the Southern hemisphere. We have argued that 
a missalignment of the dipole anisotropy with $\vec B_{IS}$ 
could be used to estimate the region of coherence of the 
local IS plasma. Although the appearance of anisotropies 
can be also understood using a diffusion equation, we think 
that our approach provides an alternative (and simpler) 
framework. In particular, Boltzmann equation averaged over different
scales provides a useful picture able to describe the changes 
in the anisotropy with the energy.

We have also suggested a mechanism that would relate
the large and the small scale anisotropies: these would appear
as the image of the global dipole 
provided by nearby cosmic magnetic lenses (Battaner et al, 2011),
that would {\it focus} the CR flow.
Notice that the lens acts as a CR source, but that the {\it real} 
source would be the large scale-anisotropy. In particular,
if this is $O(0.1\%)$, then the low-scale anisotropy will
be of the same order. If the lens is seen from the Earth under
a sizeable solid angle, the magnetic field $B_{IS}$
can define linear structures like the ones described in 
\cite{ARGO-YBJ:2013gya}. 

The simplified scheme proposed here uses a number of
approximations: all cosmic rays are protons (heavier nuclei
of the same energy would have smaller $r_L$), all cosmic rays in
the same data set have equal energy, or the effect of the
heliosphere \cite{Lazarian:2010sq} is negligible.
We think, however, that it provides
an acceptable qualitative description of the data. 
In the near future
HAWC observations from the northern hemisphere could confirm that the
TIBET/MILAGRO dipole is modulated and changes sign at 
energies above 100 TeV.

\section*{Acknowledgments}

This work has
been supported by MICINN of Spain (AYA2011-24728, 
FPA2010-16802, FPA2013-47836, 
Consolider-Ingenio {\bf Multidark} CSD2009-00064,  
{\bf EPI} CSD2010-00064), 
and by Junta de Andaluc\'\i a (FQM101, FQM108, FQM3048).


\begin{thebibliography}{99}

\bibitem{Reeves:1970zz}
  H.~Reeves, W.~A.~Fowler and F.~Hoyle,
  Nature {\bf 226} (1970) 727.

\bibitem{shalchi}
A. Shalchi, {\it Nonliner Cosmic Ray Diffusion Theories}, 
Springer-Verlag, 2009.

\bibitem{jones90}
F.C.~Jones, Astrophys.\ J.\ {\bf 361} (1990) 162.

\bibitem{Battaner:2009zf}
  E.~Battaner, J.~Castellano and M.~Masip,
  Astrophys.\ J.\  {\bf 703} (2009) L90
  [arXiv:0907.2889].

\bibitem{Ahlers:2013ima}
  M.~Ahlers,
  Phys.\ Rev.\ Lett.\  {\bf 112} (2014) 021101
  [arXiv:1310.5712].

\bibitem{Amenomori:2005dy}
  M.~Amenomori {\it et al.}  [Tibet AS Gamma Collaboration],
  Astrophys.\ J.\  {\bf 626} (2005) L29
  [astro-ph/0505114].

\bibitem{Abdo:2008aw}
  A.~A.~Abdo, B.~T.~Allen, T.~Aune, D.~Berley, S.~Casanova, 
C.~Chen, B.~L.~Dingus and R.~W.~Ellsworth {\it et al.},
  Astrophys.\ J.\  {\bf 698} (2009) 2121
  [arXiv:0806.2293]

\bibitem{ARGO-YBJ:2013gya} 
  B.~Bartoli {\it et al.}  [ARGO-YBJ Collaboration], 
  Phys.\ Rev.\ D {\bf 88} (2013) 8,  082001 
  [arXiv:1309.6182]. 

\bibitem{Guillian:2005wp}
  G.~Guillian {\it et al.}  [Super-Kamiokande Collaboration],
  Phys.\ Rev.\ D {\bf 75} (2007) 062003
  [astro-ph/0508468].

\bibitem{Mangano:2009ee}
  S.~Mangano [ANTARES Collaboration],
  ``Skymap for atmospheric muons at TeV energies measured in deep-sea neutrino telescope ANTARES,'' Proceedings of the 31st ICRC, Lodz, 2009
  [arXiv:0908.0858].

\bibitem{Aartsen:2012ma}
  M.~G.~Aartsen {\it et al.}  [IceCube Collaboration],
  Astrophys.\ J.\  {\bf 765} (2013) 55
  [arXiv:1210.5278].

\bibitem{Santander:2013gma}
  M.~Santander [IceCube Collaboration],
  Nucl.\ Instrum.\ Meth.\ A {\bf 725} (2013) 85.

\bibitem{Abeysekara:2013rka}
  A.~U.~Abeysekara {\it et al.}  [HAWC Collaboration],
  ``The HAWC Gamma-Ray Observatory: Observations of Cosmic Rays,'' 
in {\it 33rd International Cosmic Ray Conference}, Rio de Janeiro, 2013
  [arXiv:1310.0072].

 \bibitem{Zotov:2012se}
  M.~Y.~.Zotov and G.~V.~Kulikov,
  Astron.\ Lett.\  {\bf 38} (2012) 731
  [arXiv:1203.2472].
 
\bibitem{Iuppa:2013gha} 
  R.~Iuppa, 
  Frascati Phys.\ Ser.\  {\bf 55} (2012) 61 
  [arXiv:1302.7184]. 

\bibitem{schwadron2014}
N.A.~Schwadron {\it et al.}, Science {\bf 343} (2014) 988.

\bibitem{Ratkiewicz2008}
R.~Ratkiewicz and L.~Ben-Jaffel, and J.~Grygorczuk
``What do we know
about the orientation of the local interstellar magnetic field?,''
Astron.Soc.
Pac. Conf. Ser., 385, 189 – 194.

\bibitem{Frisch:2012zj}
  P.~C.~Frisch, B-GAndersson, A.~Berdyugin, V.~Piirola, R.~DeMajistre, H.~O.~Funsten, A.~M.~Magalhaes and D.~B.~Seriacopi {\it et al.},
  Astrophys.\ J.\  {\bf 760} (2012) 106
  [arXiv:1206.1273].

\bibitem{beck2005}
  R.~Beck in {\it Cosmic Magnetic Fields}, Springer Verlag, Heidelberg, 
  2005. Edited by R.~Wielebinski and R.~Beck.

\bibitem{wiel2005}
  R.~Wielebinski in {\it Cosmic Magnetic Fields}, 
  Springer Verlag, Heidelberg, 2005.
  Edited by R.~Wielebinski and R.~Beck.

\bibitem{Han:2009ts}
  J.~L.~Han,
  ``Magnetic structure of our Galaxy: A review of observations,''
 Proceedings of the IAU, Vol. 4, S259, p455 (2009) [arXiv:0901.1165].

\bibitem{edu2009}
  E.~Battaner in {\it Lecture notes and essays in Astrophysics III},
  T\'orculo, La Coru\~na, 2009. Edited by A.~Ulla and M.~Manteiga.

\bibitem{RuizGranados:2010hf}
  B.~Ruiz-Granados, J.~A.~Rubino-Martin and E.~Battaner, 
  Astronomy and Astrophysics {\bf 522} (2010) A73
  [arXiv:1006.5573].

\bibitem{Biermann:2012tc}
  P.~L.~Biermann, J.~K.~Becker, E.~-S.~Seo and M.~Mandelartz,
 Astrophys.\ J.\  {\bf 768} (2013) 124
  [arXiv:1206.0828].

\bibitem{Giacinti:2011mz} 
  G.~Giacinti and G.~Sigl, 
  Phys.\ Rev.\ Lett.\  {\bf 109} (2012) 071101 
  [arXiv:1111.2536]. 

\bibitem{Battaner:2010bd}
  E.~Battaner, J.~Castellano and M.~Masip,
  Astron.\ Astrophys.\ {\bf 527} (2011) 5 
  [arXiv:1006.2346].

\bibitem{Pohl:2012} 
M.~Pohl and D.~Eichler, 
  Astrophys.\ J.\  {\bf 766} (2013) 4
[arXiv:1208.5338] 

\bibitem{Qu:2011fe}
  X.~-B.~Qu, Y.~Zhang, L.~Xue, C.~Liu and H.~-B.~Hu,
  Astrophys.\ J.\  {\bf 750} (2012) L17
  [arXiv:1101.5273].

\bibitem{McComas:2012}
  D.~J.~McComas, D.~Alexashov, G.~Bzowski {\it et al.},
  Science.\ Vol.\ {\bf 336} (2012) 1221054.

\bibitem{C&G}
  A.H.~Compton and I.A.~Getting,
  Phys.\ Rev.\ {\bf 47} (1935) 817.

\bibitem{Swordy}
S.~P.~Swordy,
Space Science Reviews 99 (2001) 85.

\bibitem{Salvati:2010pq} 
  M.~Salvati, 
  Astron.\ Astrophys.\ {\bf 513} (2010) A28
  [arXiv:1001.4947]. 

\bibitem{Lemaître:1934}
 G.~Lemaître and M.~Vallarta, 
 Phys.\ Rev.\, 43 (1933) 89.

\bibitem{DiSciascio:2011fk}
  G.~Di Sciascio {\it et al.}  (2012) [ARGO-YBJ Collaboration],
  J.\ Phys.\ Conf.\ Ser.\  {\bf 375} 052008
  [arXiv:1112.0666 [astro-ph.HE]].

\bibitem{Amenomori:2006bx}
  M.~Amenomori {\it et al.} (2006) [Tibet AS-gamma Collaboration],
  Science {\bf 314} 439
  [astro-ph/0610671].

\bibitem{Abbasi:2011zka}
  R.~Abbasi {\it et al.}  (2012) [IceCube Collaboration],
  Astrophys.\ J.\  {\bf 746} 33
  [arXiv:1109.1017 [hep-ex]].

\bibitem{Aglietta:2009mu}
  M.~Aglietta {\it et al.} (2009) [EAS-TOP Collaboration],
  Astrophys.\ J.\  {\bf 692} L130
  [arXiv:0901.2740 [astro-ph.HE]].

\bibitem{Casse:2001be}
  F.~Casse, M.~Lemoine and G.~Pelletier,
  Phys.\ Rev.\  D {\bf 65} (2002) 023002
  [arXiv:astro-ph/0109223].

\bibitem{Blasi:2011fm}
  P.~Blasi and E.~Amato,
  JCAP {\bf 1201} (2012) 011
  [arXiv:1105.4529].

\bibitem{Abdo:2008kr}
  A.~A.~Abdo, {\it et al.}  (2008), 
  Phys.\ Rev.\ Lett.\  {\bf 101} 221101
  [arXiv:0801.3827 [astro-ph]].

\bibitem{Lazarian:2010sq}
  A.~Lazarian and P.~Desiati,
  Astrophys.\ J.\  {\bf 722} (2010) 188
  [arXiv:1008.1981].


\end{thebibliography}
\end{document}